\documentclass[twocolumn]{aastex7}
\begin{document}
\title{The Role of Faraday Rotation in the Polarization of the X-rays 
from Magnetically Powered Black Hole Coronas}

\author[orcid=0000-0002-1084-6507,gname='Henric',sname='Krawczynski']{Henric Krawczynski}
\affiliation{Washington University in St. Louis, Physics Department, the McDonnell Center for the Space Sciences, and the Center for Quantum Leaps, St. Louis, MO 63130}
\email[show]{krawcz@wustl.edu}  

\begin{abstract}
Magnetic reconnection is one of the prime candidate mechanisms that may 
energize the plasma emitting the strongly polarized X-ray emission from
black hole X-ray binaries (BHXRBs) in their hard states.
The mechanism requires strong magnetic fields in the 
upstream plasma entering the reconnection layer, and weaker, but 
still substantial, magnetic fields 
in the downstream regions.
In this Letter, we estimate the coronal magnetic fields for three 
different magnetic energy dissipation mechanisms: plasmoid-dominated 
magnetic reconnection, fast collisionless reconnection, and magnetic field relaxation. 
We show that the lack of strong Faraday depolarization constrains viable models and can be used to benchmark numerical accretion flow models. We conclude by discussing the difficulties of disentangling the various effects that can depolarize the signals from BHXRBs at low energies. We furthermore emphasize that Faraday rotation is unlikely to play a role in the polarization 
of the coronal X-ray emission 
of active galactic nuclei.
\end{abstract}
\section{Introduction}
The Imaging X-ray Polarimetry Explorer (IXPE) measured the soft-state
and hard-state polarization of 
black hole X-ray binaries
(BHXRBs) over 
the 2-8 keV energy band \citep{2022cosp...44.1863R,2024cosp...45.1615I,2024Galax..12...54D,2026MNRAS.545f1933M}.
The observations included sources emitting in the soft state and the hard state.
The soft-state emission is attributed to the multitemperature blackbody emission from a 
geometrically thin, optically thick accretion disk \citep{1973A&A....24..337S,1973blho.conf..343N,2019ApJ...874...23D,2022mgm..conf..250T}.
The hard-state emission comes from the Comptonization of longer-wavelength seed photons in hot plasma \citep{1976ApJ...204..187S,1976ApJ...206..910K,
1980A&A....86..121S,1979A&A....75..214P,1995ApJ...450..876T}. Alternatively, the photons may Comptonize by repeated scatterings off relativistically moving plasmoids  or flux ropes \citep{2017ApJ...850..141B,LS2016,2020ApJ...899...52S,
2021MNRAS.507.5625S,
2023MNRAS.518.1301S,
2025ApJ...979..199S}.

This Letter discusses the impact of Faraday rotation on the polarization of the X-ray emission from BHXRBs. Several authors discussed the Faraday rotation of the emission from geometrically thin, optically thick accretion disks. \citet{1996MNRAS.282..965A} found that a magnetic field in equipartition with the radiation from the accretion disk emission can strongly modify the polarization of the optical emission from the accretion disks 
of active galactic nuclei (AGNs). \citet{1998MNRAS.293....1A} and \citet{2006AstL...32...39G} presented refined estimates accounting for absorption and scattering processes 
in the accretion disk atmosphere.  
\citet{2009ApJ...703..569D} discussed Faraday rotation of the X-rays from BHXRBs in the soft state.
Using the results from radiation magnetohydrodynamic simulations of the local vertical accretion disk structure, they found that ``the magnetic fields in the simulations are just strong enough to produce significant Faraday depolarization near the spectral peak of the radiation field,'' i.e., at energies around 1\,keV.
Faraday depolarization refers here to the effect by which varying amounts of Faraday rotation for different portions of the signal weaken the overall net polarization.
The analyses of the IXPE BHXRB results in the soft state have so far largely ignored 
this result, although some soft-state observations, e.g., those of Cyg X-1   \citep{2024ApJ...969L..30S}, revealed marked polarization degree (PD) increases with energy that 
may indeed result from the stronger Faraday depolarization at lower energies than at higher energies.

In this Letter we discuss the Faraday depolarization of the hard-state emission from BHXRBs. 
The IXPE observations revealed unexpectedly high PDs. The source Cyg X-1 showed PDs of $\sim$4\% -- surprisingly high for a low-inclination system \citep{2022Sci...378..650K,2025A&A...701A.115K}.
The PDs of the hard-state emission of the source Swift~J1727.8$-$1613 were found 
to be around $\sim$4\% as well  \citep{2023ApJ...958L..16V,2024A&A...686L..12P,2024ApJ...968...76I,2024RAA....24i5004L}.
The source IGR~J17091$-$3624 
exhibits extremely high PDs of 
about 10\%
\citep{2025MNRAS.541.1774E,2025ApJ...989..165D}.
If the corona is compact and has a size of a few $r_{\rm g}$ 
($r_{\rm g}\,=\,GM/c^2$ is the gravitational radius of the black hole with mass $M$, and $G$ and $c$ are the gravitational constant and the speed of light, respectively), hot 100\,keV plasma radiatively cools on timescales shorter 
than the light travel time across the corona \citep{1979ApJ...229..318G,2017ApJ...850..141B,2025ApJ...993...54K}. 
Compact coronas thus need to be heated 
in situ. Consequently, various authors  have argued that the energy transport into the corona is dominated by magnetic fields \citep[e.g.,][]{1979ApJ...229..318G,2008ApJ...688..555G,2017ApJ...850..141B}.

The IXPE results indicate that Faraday depolarization does not play a major role in BHXRBs in the hard state. In the rest of this Letter, we discuss this result in the context of magnetically dominated corona models.
We begin in Section\,\ref{ml} with a discussion of the mechanisms that can convert magnetic energy into plasma bulk motion and heat. In Section\,\ref{results},  we use the observed luminosity of 
the Cyg X-1 to set lower limits on the coronal magnetic field strength. 
Combining these limits with estimates of the electron (and possibly positron) densities in the corona, 
we derive the expected Faraday depolarization.
We end with a summary and discussion of 
the results in Section \ref{discussion}.
\section{Magnetoluminescence Mechanisms}
\label{ml}
%
In the following we parameterize the mechanism of magnetoluminescence (conversion of magnetic field energy into radiation) with two parameters: the speed $\beta_{\rm dis}$ with which the magnetized plasma moves into the dissipation region in units of the speed of light $c$, and the efficiency $\eta_{\rm rad}$ with which the magnetic energy is converted into radiation.
As detailed below, $\beta_{\rm dis}$ depends on the dissipation mechanism, and we use values of  between 0.01 and 1. 
The efficiency $\eta_{\rm rad}$ depends on many different factors, including the fraction of the magnetized plasma that participates in the energy conversion, and the efficiencies of the conversion of magnetic energy into 
kinetic energy and heat, and from these two forms of energy, into radiation.

Classical Sweet--Parker reconnection proceeds too slowly to explain the observed BHXRB luminosities \citep[e.g.,][]{2005ppa..book.....K}.  
If the plasma is collisionless, reconnection can proceed much faster, with the magnetized plasma entering the reconnection region at a tenth of the speed of light  \citep[$\beta_{\rm dis}\,\sim\,0.1$;][]{2006ApJ...644L.145C,2010PhRvL.105w5002U,2016ASSL..427..473U,2025ARA&A..63..127S}.  
Fast reconnection is possible if the Sweet--Parker reconnection layer 
thickness $\delta_{\rm SP}$ is smaller than 
the ion skin depth $d_{\rm i}$ below 
which the plasma ceases to behave 
as a single fluid. \citet{2008ApJ...688..555G} 
analyze the physical conditions in 
BHXRB (and AGN) coronas and show that \begin{equation}
\frac{\delta_{\text{SP}}}{d_{\rm i}} \approx \sqrt{\frac{m_{\rm e}}{m_{\rm p}}} \ln \Lambda f^{-1/4} \dot{m}^{-1/4} \tau^{3/4} \vartheta_{\rm e}^{-3/4} (2\,\rho)^{3/8} 
\label{e:ratio}
\end{equation}
when Coulomb interactions dominate the magnetic diffusivity.
Here, $f$ is the fraction of the angular momentum of the accreted matter being carried away by the corona, $m_{\rm e}$ and $m_{\rm p}$ are the electron mass and proton mass, respectively, $\ln \Lambda\,\sim\,20$ is the Coulomb logarithm, $\dot m$ is the mass accretion rate in units of $L_{\rm Edd}/c^2$  
(with $L_{\rm Edd}$ being the Eddington luminosity),  
$\tau$ is the optical depth of the coronal plasma,
$\vartheta_{\rm e}$ is the coronal temperature of the plasma in units of $m_{\rm e} c^2/k_{\rm B}$, and 
$\rho$ is the distance from the black hole in units of $r_{\rm g}$.
Using the following values for 
the hard state of Cyg X-1: 
$f\,=\,1$,
$\dot{m}\,=\,1$,
$\tau\,=\,1$, and
$\vartheta_{e}\,=\,0.2$
\citep{2025ApJ...993...54K},
we infer
\begin{equation}
\frac{\delta_{\rm SP}}{d_{\rm i}}\approx  0.45 \, \rho^{3/8}
\end{equation}
indicating that the condition 
$\delta_{\rm SP}/d_{\rm i}\,<\,1$ can be met for
compact coronas at radial distances $r\,\lesssim\,8\,r_{\rm g}$.
The authors emphasize that inverse Compton processes increase the collisionality 
of the plasma close to the black hole -- but in a subdominant way. 

Reconnection may proceed much faster than in the Sweet--Parker case -- even if the collisionless condition is not met. For large Lundquist numbers $v_A\,L/\eta^>_{\sim}10^4$ with the Alfv\'en speed $v_A\,\sim\,c$ for a highly magnetized plasma, $L$ the length of the current sheet, and $\eta$ the resistivity, 
plasmoid instabilities
can lead to a largely 
$\eta$-independent reconnection rate with the plasma streaming into the reconnection zone with $\beta_{\rm dis}\,=\,0.01$
\citep{2009PhPl...16k2102B,2010PhRvL.105w5002U,2020ApJ...900..100R}.
Recent particle-in-cell simulations and relativistic resistive magnetohydrodynamic simulations with non-uniform resistivities indicate that the reconnection velocity can reach $\beta_{\rm dis}\sim 0.1$ \citep[][and references therein]{2026arXiv260102460R}.

A third possibility of converting magnetic energy 
into kinetic energy involves magnetic field relaxation, which preserves the magnetic field line topology \citep{2017SSRv..207..291B}. Magnetic energy can, for example, be released if slip knots resolve
and magnetic pressure and tension are released. 
In this case, the energy can move with $\beta_{\rm dis}=1$ into the dissipation region.

\section{Expected Faraday Rotation in the IXPE Energy Band}
\label{results}
In the following, we derive estimates of the magnetic field in the coronal plasma. 
We give all estimates for the BHXRB Cyg X-1, which has well-defined system parameters.
The binary harbors an $M\,=\,21.2\pm2.2\,M_{\odot}$ black hole orbiting 
an O-star of mass 40.6$^{+7.7}_{-7.1}\,M_{\odot}$ in a 5.599829(16) day orbit. 
The binary system is seen 
at an inclination of $i\,=\,27^{\circ}.51^{+0.77}_{-0.57}$
from its orbital axis \citep{2021Sci...371.1046M,2008ApJ...678.1237G}.  
The semimajor axis $a_{\rm bin}$ 
of 0.244 au is 2.35 times larger than 
the radius $R_1$ of the companion 
star of 22.3$^{+1.8}_{-1.7}$
$R_{\odot}$. The eccentricity of the orbit is 0.0189$^{+0.0028}_{-0.0026}$.

The average hard-state bolometric luminosity is $L\,=\,0.005\,L_{\rm Edd}$ \citep{2002ApJ...578..357Z,2003MNRAS.343L..84G}.
Assuming isotropic emission and neglecting 
special relativistic effects (owing to plasma motion) and
general relativistic effects (close to the black hole), the energy density in the emission region is
\begin{equation}
U_{\text{rad}} = \frac{L}{4 \pi R^2 c}.
\end{equation}
With the assumptions from the previous section, 
we derive the magnetic energy density of the plasma
streaming from the accretion disk between radii $r_1$ and $R$ into the corona:
\begin{equation}
U_{\text{B}} = \frac{L}{2\pi(R^2-r_1^{\,2})\,\beta_{\rm dis}\,c\,\eta_{\rm rad}}
    \label{e:uB}
\end{equation}
as well as the magnetic field strength:
\begin{equation}
B = \sqrt{8\pi\,U_{\rm B}}.
    \label{e:B}
\end{equation}
Assuming a black hole with a spin parameter of 
$a=0.998$, the innermost stable circular orbit is at   $r_1\,=\,1.237\,r_{\rm g}$. 
With $L\,=\,0.5\%\,L_{\rm Edd}$, $r_1\,=\,1.237\,r_{\rm g}$, and $R\,=\,5\,r_{\rm g}$, we infer a magnetic field of
\begin{equation}
    B\,=\,4.3\times 10^7 \,{\rm G}\,
\left(
\frac{\beta_{\rm dis}}{0.01}\cdot
\frac{\eta_{\rm rad}}{0.5}
\right)^{-1/2}
    \label{e:B2}
\end{equation}
This gives magnetic fields of
(i) 4.3$\times 10^7$\,G for plasmoid-dominated reconnection, 
(ii) 1.35$\times 10^7$\,G for 
fast collisionless reconnection, and 
(iii) 4.3$\times 10^6$\,G for magnetic field relaxation.
The observed polarization of an X-ray from the corona largely originates in the last scattering 
with previous scatterings impacting the orientation of the scattering plane. 
The Faraday rotation angle is thus proportional to the line integral of the magnetic 
field along the photon path from the location of the last scattering to the location 
where the photon leaves the corona:
\begin{equation}
\Delta\chi_0 = 
\frac{3\sigma_{\rm T}}{16\pi^2\,e}\lambda^2
\int n_{\rm e} \mathbf{B} \cdot d\mathbf{s}.
\label{e:0}
\end{equation}
IXPE observes BHXRBs for between 10 and a few hundred kiloseconds, much longer than
$r_{\rm g}/c \sim 0.1$\,ms. The observations thus likely average over many
separate flares from different regions with changing  magnetic field configurations 
and Faraday rotation angles. The net effect is a depolarization that depends 
on the rms of the Faraday rotation angle. 
For randomly oriented magnetic fields, the rms of $\Delta\chi$ is $\Delta\chi_0/\sqrt{3}$. 

If the magnetic field changes direction $N_{\rm reg}$ times between the last scattering and the photon escaping the emission region, $\Delta\chi$ performs a random walk, reducing the expected net rms by $\sqrt{N_{\rm reg}}$, so that we get
\begin{equation}
{\rm rms}(\Delta\chi)\,=\,\frac{\Delta\chi_0}{\sqrt{3\,N_{\rm reg}}}.
\label{e:rms1}
\end{equation}
This rms neglects the variations of the Faraday rotation owing to the statistical variations of 
the depth of the last scattering. 

Assuming that the corona has a height $h\,r$ at the radial 
distance $r$ 
from the black hole and the corona presents an optical depth $\tau$ to photons originating 
from the accretion disk, we infer an electron density of
\begin{equation}
n_{\text{e}} = \frac{\tau} {h\,r\,\sigma_{\text{T}}}. 
\label{e:ne}
\end{equation}
If a photon travels a distance of 
\begin{equation}
\Delta s\,\sim\, \frac{h\,r}{\tau}
\label{e:ds}
\end{equation}
between the last scattering and the escape from the corona, we can combine Equations (\ref{e:0})-(\ref{e:ds}) to get
\begin{equation}
{\rm rms}(\Delta\chi)\,=\,
\frac{\sqrt{3}\,\lambda^2\,B}{16\pi^2\,e\,\sqrt{N_{\rm reg}}}.
\label{e:rms2}
\end{equation}
The equation is only valid for $\tau^>_{\sim}1$ as the assumption $\Delta s\,\sim\, hr/\tau$ breaks down otherwise. For $\tau\,<\,1$ the result needs to be scaled with $\tau$.

Assuming that the X-rays propagate through the magnetic field from Equation (\ref{e:B2}), we infer
\begin{equation}
{\rm rms}(\Delta\chi)=
3.74\,{\rm rad}
\left(\frac{E}{2\,\rm keV}\right)^{-2}\!
\left(
\frac{\beta_{\rm dis}}{0.01}\!\cdot\!
\frac{\eta_{\rm rad}}{0.5}\!\cdot\!
N_{\rm reg}
\right)^{-1/2}.
\label{e:rms3}
\end{equation}
Some of the plasma through which the X-rays propagate may be less magnetized (e.g., in the downstream region) than given by Equation (\ref{e:B2}), but the estimate should still be correct to leading order. 
Using $N_{\rm reg}\,=\,1$ for the time being, we thus estimate 2\,keV rms Faraday rotations of (i) 3.74\,rad for plasmoid-dominated reconnection, (ii) 1.18\,rad for 
fast collisionless reconnection, and (iii) 0.37\,rad for magnetic field relaxation.

\begin{figure}
  \centering
  \includegraphics[width=0.45\textwidth]{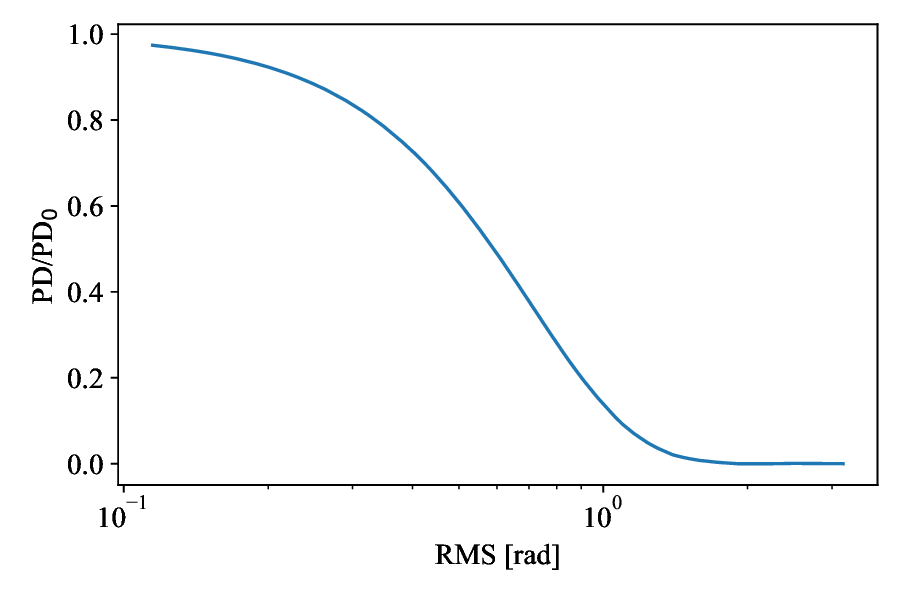}
  \caption{Ratio of the polarization degree of a signal after and before Faraday depolarization as a function of the rms of Faraday rotation.}
  \label{f:depol}
\end{figure}
Figure \ref{f:depol} shows the depolarization from Faraday rotation 
as a function of the rms from a toy simulation. 
An rms of 0.59 rad (34$^{\circ}$) lowers the PD by 50\%. 
The PD drops quickly to $<$1\% of the original value for rms values equal or larger than $\pi/2$. 
Thus, rms-values exceeding $\pi/2$ are  not compatible with the observations of strongly 
polarized BHXRB signals.
Reducing the original polarization by less than 10\% requires rms-values smaller than 0.22 rad. This condition is met if $(\beta_{\rm dis}/0.01)(\eta_{\rm rad}/0.5)N_{\rm reg}>265$.
\section{Discussion}
\label{discussion}
The IXPE detections of highly polarized signals from the 
BHXRBs Cyg~X$-$1, Swift~J1727.8$-$1613, and IGR~J17091$-$3624 
in the hard state show that Faraday depolarization does not entirely suppress the polarization signals. This is in some tension with our estimates of the Faraday depolarization in magnetic field dominated coronas. The model with the lowest upstream plasma velocity, 
plasmoid-dominated reconnection, requires the strongest 
magnetic fields, and predicts the highest Faraday depolarization. 
Fast reconnection and magnetic field relaxation predict 
lower degrees of Faraday depolarization. 
Major uncertainties in the estimates of the Faraday depolarization 
stem from the unknown structure of the magnetic field in the 
emission regions and their surroundings, as well as from the
theoretically and observationally poorly constrained overall 
geometry of the accretion flow in the hard state. 
The following effects may impact the degree of Faraday depolarization: 
\begin{enumerate}
\item If the magnetic field direction reverses along the photon path after the last scattering several times and $N_{\rm reg}> 1$, Faraday rotation is suppressed by a factor of $\sqrt{N_{\rm reg}}$. 
\citet{2025ApJ...979..199S} present global resistive general relativistic magnetohydrodynamic 
simulations showing the magnetic field in the sheath between the accretion flow 
and the jet. The magnetic field topology inside the sheath (their Fig. 5) exhibits a few 
field reversals, indicating $N_{\rm reg}$ may be on the order of a few. 
\citet{2024NatCo..15.7026N} and \citet{2024PhRvL.132h5202G} use 3D particle-in-cell simulations to simulate turbulent plasma. The turbulence creates a highly tangled magnetic field with $N_{\rm reg}\gg 1$. 
Note that reconnection in 3-D likely always creates turbulence.
\item For a given magnetoluminescence mechanism and fixed $\beta_{\rm dis}$,
the requirement rms$(\Delta\chi)\ll\pi/2$ constrains the product 
$\eta_{\rm dis}\times N_{\rm reg}$. Note that $\eta_{\rm dis}\ll1$ 
seems quite likely, given that the geometry of the field lines will not always be favorable for highly efficient dissipation. It would be interesting to measure the product $\eta_{\rm dis}\times N_{\rm reg}$ 
in numerical accretion flow simulations to test if values of order unity or larger can be achieved. The requirements 
of high $\eta_{\rm rad}$ 
and a scrambled magnetic field
may be mutually exclusive.
The IXPE observations of BHXRBs in the hard state can thus be 
used to benchmark numerical accretion flow simulations.
\item A possible outcome of reconnection, particle acceleration, and $e^+e^-$ pair creation may be that most of the particles end up inside plasmoids and that the regions between plasmoids are tenuous \citep{2024MNRAS.52711587M,2025ARA&A..63..127S}. The plasmoids can thus have  
high density producing the required scattering optical depth. The low particle 
density between plasmoids can lead to less Faraday rotation than predicted by 
Equation (\ref{e:rms3}). 
\item The coronal region may be substantially larger than $R=5\,r_{\rm g}$, reducing the magnetic field strength required to power the coronal emission (Equations (\ref{e:uB}) and (\ref{e:B})).
For a 20 times larger corona ($R\,=100\,r_{\rm g}$), the magnetic field and the Faraday rotation angle are by a factor of 20 smaller, 
moving the energy band of significant depolarization to energies well below the IXPE energy band.
The corona is unlikely to be much larger than this given 
luminosity constraints \citep{2025ApJ...993...54K} and the limits on the corona size 
from observed flare durations \citep{2003MNRAS.343L..84G}.
\end{enumerate}

Has IXPE already observed Faraday depolarization in the PD energy spectra of BHXRBs? Owing to the symmetry of the inner accretion flow onto black holes, averaging over partially depolarized beams with different amounts of Faraday rotation 
is likely to produce a net polarization angle parallel or perpendicular to the angular momentum vector of the inner accretion disk (thought to be aligned with the black hole spin axis).
The effect of Faraday rotation would thus likely be observed as a depolarization at low energies. 
The PD energy spectra of Cyg X-1 and Swift J1727.8$-$1613 indeed show
PDs increasing with energy. 
A fraction of this increase may indeed come from the Faraday depolarization at lower energies.
Note, however, that several effects other than Faraday rotation may contribute to the low PDs 
at low energies, i.e., the competition between the polarization of the disk emission, 
the coronal emission, and the reflected strongly gravitationally lensed emission
\citep{2009ApJ...701.1175S,2010ApJ...712..908S,2022ApJ...934....4K}. 
Disentangling these effects can thus be rather challenging in practice.

Faraday depolarization should be stronger at energies below IXPE's 2-8 keV band, and should be less significant at higher energies.
Broadband X-ray polarization observations with missions such as the
Rocket Experiment Demonstration of a Soft X-ray Polarimeter
or the Globe-Orbiting Soft X-ray Polarimeter in the 0.2-0.4 keV energy band
\citep{2025SPIE13625E..10T,2022cosp...44.1871M}
or with an XL-Calibur-type space mission covering the 3-60 keV energy band
\citep{2016APh....75....8K,2021APh...12602529A,2025ApJ...994...37A}
would help to disentangle the different effects.

How much does Faraday rotation affect the coronal X-ray emission from AGNs?
Assuming identical luminosities in units of the Eddington luminosity, 
and identical corona sizes in units of $r_{\rm g}$, the magnetic field and 
the rms of the Faraday rotation angles scale as $1/\sqrt{M}$ 
with the black hole mass $M$ (see Equations (\ref{e:uB}) and (\ref{e:B})).
For an AGN with a $M\,=\,10^8 M_{\odot}$ black hole, the expected rms of 
the Faraday rotation angles is thus by a factor of $\sim 2000$ smaller than for Cyg X-1, and is thus negligible. 
\begin{acknowledgments}
The author thanks Yajie Yuan, Alex Chen, Kun Hu, John Groger,
Shravan Vengalil Menon, Daniel Gro{\v{s}}elj, Andrei Beloborodov, Lorenzo Sironi, Navin Sridhar, Bart Ripperda, 
Matthew Liska, Bert Vander Meulen for joint discussions. 
He furthermore acknowledges NASA support with the grants 80NSSC24K1178, 80NSSC24K1749, and 80NSSC24K1819, as well as support from the McDonnell Center for the Space Sciences 
at Washington University in St. Louis.
\end{acknowledgments}
\bibliography{manuscript}{}
\bibliographystyle{aasjournalv7}
\end{document}